\documentclass[aps,pre,floats,twocolumn,showpacs,superscriptaddress]{revtex4}

\usepackage{graphicx,epsfig}
\usepackage{times}
\usepackage{graphics,dcolumn,bm,fleqn,epic,eepic,float}
\usepackage{amssymb,amsmath,multirow,rotate,color}
\begin{document}

\title{Dynamical organization towards consensus in the Axelrod model
  on complex networks}

\author{Beniamino Guerra}

\affiliation{Laboratorio sui Sistemi Complessi, Scuola Superiore di Catania, 
Via San Nullo 5/i, 95123 Catania, Italy} 

\author{Julia Poncela}

\affiliation{Institute for Biocomputation and Physics of Complex
Systems (BIFI), University of Zaragoza, 50009 Zaragoza, Spain}

\author{Jes\'us G\'omez-Garde\~nes}
\email{gardenes@gmail.com}
\affiliation{Institute for Biocomputation and Physics of Complex
Systems (BIFI), University of Zaragoza, 50009 Zaragoza, Spain}

\affiliation{Departamento de Matem\'atica Aplicada, Universidad Rey Juan Carlos (ESCET), 95123 M\'ostoles (Madrid), Spain}

\author{Vito Latora}

\affiliation{Dipartimento di Fisica e Astronomia, Universit\`a di
  Catania and INFN, Via S. Sofia, 64, 95123 Catania, Italy}

\affiliation{Laboratorio sui Sistemi Complessi, Scuola Superiore di Catania, 
Via San Nullo 5/i, 95123 Catania, Italy} 

\author{Yamir Moreno}

\affiliation{Institute for Biocomputation and Physics of Complex
Systems (BIFI), University of Zaragoza, 50009 Zaragoza, Spain}

\affiliation{Department of Theoretical Physics, University of Zaragoza, 50009 Zaragoza, Spain}

\date{\today}
\begin{abstract}
  We analyze the dynamics towards cultural consensus in the Axelrod
  model on scale-free networks. By looking at the microscopic dynamics
  of the model, we are able to show how culture traits spread across
  different cultural features. We compare the diffusion at the level
  of cultural features to the growth of cultural consensus at the
  global level, finding important differences between these two
  processes. In particular, we show that even when most of the
  cultural features have reached macroscopic consensus, there are
  still no signals of globalization. Finally, we analyze the topology
  of consensus clusters both for global culture and at the feature
  level of representation.
\end{abstract}

\maketitle

\section{Introduction}


The study of social systems has attracted the interest of statistical
physics in the recent years. In particular, this interest relies in
the description of the global behaviors that emerge in the social
context via simple models incorporating local interactions between
individuals. Examples of these global behaviors include culture
dissemination, the emergence of cooperation, and the formation of
opinions (see \cite{sociophysics} for a recent review). Although
most social models developed are usually casted as too simple from the
point of view of the interaction rules, the statistical physics
approach has successfully captured the essential features of emerging
social behaviors, showing that microscopic details of the local
processes are not relevant to explain the macroscopic emergent
phenomena.

Additionally, current approaches also incorporate the topology of
interactions as a key ingredient to describe social dynamics. Although
traditional statistical physics is usually concerned with ordered and
regular structures, such as lattices, and mean-field approximations,
large social systems are better described by complex networks of
interactions
\cite{Albert:2002,Dorogovtsev:2002,Newman:2003,Boccaletti:2006}. Complex
social networks show nontrivial topological properties, such as
power-law (scale-free) probability distributions for the number of
neighbors of individuals, high clustering and modularity among other
properties. These structural properties influence different dynamical
behaviors \cite{DoroRev,SyncRev} and, in particular, they have been
shown to play a key role in the emergence of many social phenomena
such as the enhancement of cooperation in populations interacting
within an evolutionary context \cite{Szabo}.



Here we will focus on the dissemination of culture in social
networks. There are many different models that describe how
individuals are culturally influenced by their local neighborhood
\cite{sociophysics}, however, most of the models rely on the principle
that individuals tend to become more culturally alike when they
interact. In this context, the social phenomenon to be described is
how globalization, {\em i.e.} the state in which nearly all the system
reaches a cultural consensus, emerges, in contrast with a state of
cultural fragmentation where diverse cultural groups coexist. A famous
model of cultural dissemination was introduced by Robert Axelrod
\cite{Axelrod:1997,Axelrod-2}. In this model the culture of each
individual is represented by a vector whose components are the {\em
  cultural features}. Each of these features can take a limited number
of values or {\em cultural traits}. The Axelrod model assumes that the
more similar are two interacting individuals, the more similar they
tend to become (homophily), and thus interactions favor the onset of
cultural consensus following the social principle stated above.  The
Axelrod model has both global cultural consensus and fragmentation
into diverse cultural clusters as possible frozen equilibria.  The
Axelrod model was originally implemented in a square lattice and it
was shown how, depending on the number of cultural features and
traits, cultural consensus or social fragmentation were obtained as
final states of the dynamics. The model was later analyzed in the
light of statistical mechanics to characterize the non-equilibrium
order-disorder (consensus-fragmentation) phase transition
\cite{Castellano00}. From this perspective, several studies have
considered the Axelrod model to study the influence of external fields
\cite{Gonzalez-Avella}, the role of dimensionality \cite{Klemm}, and
the effects of noise \cite{Klemm2} and mobility of agents
\cite{PREmob}. In addition to this, the mean-field description of the
model has been analyzed in Ref.~\cite{Castellano00,Vilone,Redner},
while the Axelrod model in complex topologies has been studied in
Ref.~\cite{Klemm3}. On a complex network, the transition between the
ordered homogeneous state and the disordered state is shown to be
shifted by the heterogeneity of the underlying topology. Namely, the
more heterogeneous the distribution of the number of contacts, the
more robust the cultural consensus phase.

In this work we study the Axelrod model in complex networks from a
different perspective. We will focus on the dynamics towards the
equilibrium configurations rather than on the statistical
characterization of the order-disorder phase transition. In
particular, we will study how the system self-organizes to reach
global consensus or cultural fragmentation and the microscopic origin
of the time-costly reorganization processes that take place before the
dynamics gets trapped in a frozen equilibrium.
Instead of characterizing, as usual, the state of the system by the
number of culturally identical individuals, we will consider how
cultural traits propagates across cultural features. By looking at
these internal processes we will show how cultural clusters are formed
inside each cultural feature and characterize their dynamics and
topology. Our results show that the dynamics inside the cultural
features differs drastically from that observed looking at the global
culture level since most of the cultural features can achieve
macroscopic consensus by their own in a fast way while in the whole
system no signals of cultural consensus are yet observed. This fast
time scale for the organization of most cultural features becomes
screened when looking at the formation of global
consensus. Additionally, our results point out that the time-costly
reorganization processes that delay the final cultural equilibrium are
localized in few cultural features rather than taking place at all the
feature levels.

\section{The Axelrod model in complex networks}

The Axelrod model is implemented on a complex topology by considering
that each individual is placed at a different node of the network. In
this way, a network of $N$ vertices represents a system of $N$
interacting individuals.  The links of the network account for the
interactions pattern of the social system, so that the agents
interacting with an individual $i$ are its first neighbors in the
graph.  In particular we consider scale-free (SF) networks,
{\em i.e.} graphs having a power law degree distribution,
$P(k)=k^{-\gamma}$. This kind of graphs, with an exponent $2
<\gamma\leq 3$, are usually observed in different complex systems,
including social ones. This finding implies that social systems are
highly heterogeneous: in a social network many individuals have a
small number of acquaintances, while a few agents are largely
connected with the rest of the population. Mathematically the
heterogeneity is clear in the thermodynamic limit,
$N\rightarrow\infty$, where the second moment of the degree
distribution, $\langle k^2\rangle$, diverges. In this work we will
focus the study on SF networks with $\gamma=3$ as obtained from the
Barab\'asi-Albert model \cite{BA}.

As introduced above, in the Axelrod model each agent is represented by
a vector ${\bf v}_i=(v_{i}^{1}, v_{i}^2, ...,v_{i}^{F})^T$, with
$i=1,...,N$, of $F$ components, the so-called {\em cultural
  features}. Each of these components can take only $Q$ integers
values, the {\em cultural traits}. We assume, as usual, that the value
$Q$ is the same for the $F$ components. Initially, we assign random
values (with equal probability $1/Q$) to each of the cultural features
of the $N$ agents in the system. At each time step, we randomly choose
one node $i$ and pick randomly one of its neighbors, say $j$, ({\em
  i.e.} we select a pair of connected agents). Then we check the
similarity between the agents $i$ and $j$. The similarity or overlap
between $i$ and $j$ is defined as:
\begin{equation}
S_{ij}=\frac{1}{F}\sum_{l=1}^{F}\delta(v_{i}^{l}-v_{j}^{l})\;,
\end{equation}
where $\delta(x)=1$ if $x=0$ and $\delta(x)=0$ otherwise. If the
individuals are totally different ($S_{ij}=0$) or they share identical
cultural traits ($S_{ij}=1$) nothing happens and we consider the link
between them as ``{\em blocked}''. On the other hand, when
$S_{ij}\in(0,1)$, the link between them is ``{\em active}'' and we
consider the similarity value $S_{ij}$ as the probability that agents
$i$ and $j$ decide to interact becoming culturally closer in the next
generation. In this way, if the agents are culturally far, the
probability that they decide sharing the same trait for one cultural
feature is very low, whereas individuals sharing many common cultural
features are very likely to become culturally identical after few
interactions. In case of interaction, we choose randomly a cultural
trait $f$ such that $v_{i}^{f}\neq v_{j}^{f}$ and we set $v_{i}^{f} =
v_{j}^{f}$, {\em i.e.}  $i$ copies the culture trait $v_{j}^f$ of $j$.

\begin{figure}[!t]
\epsfig{file=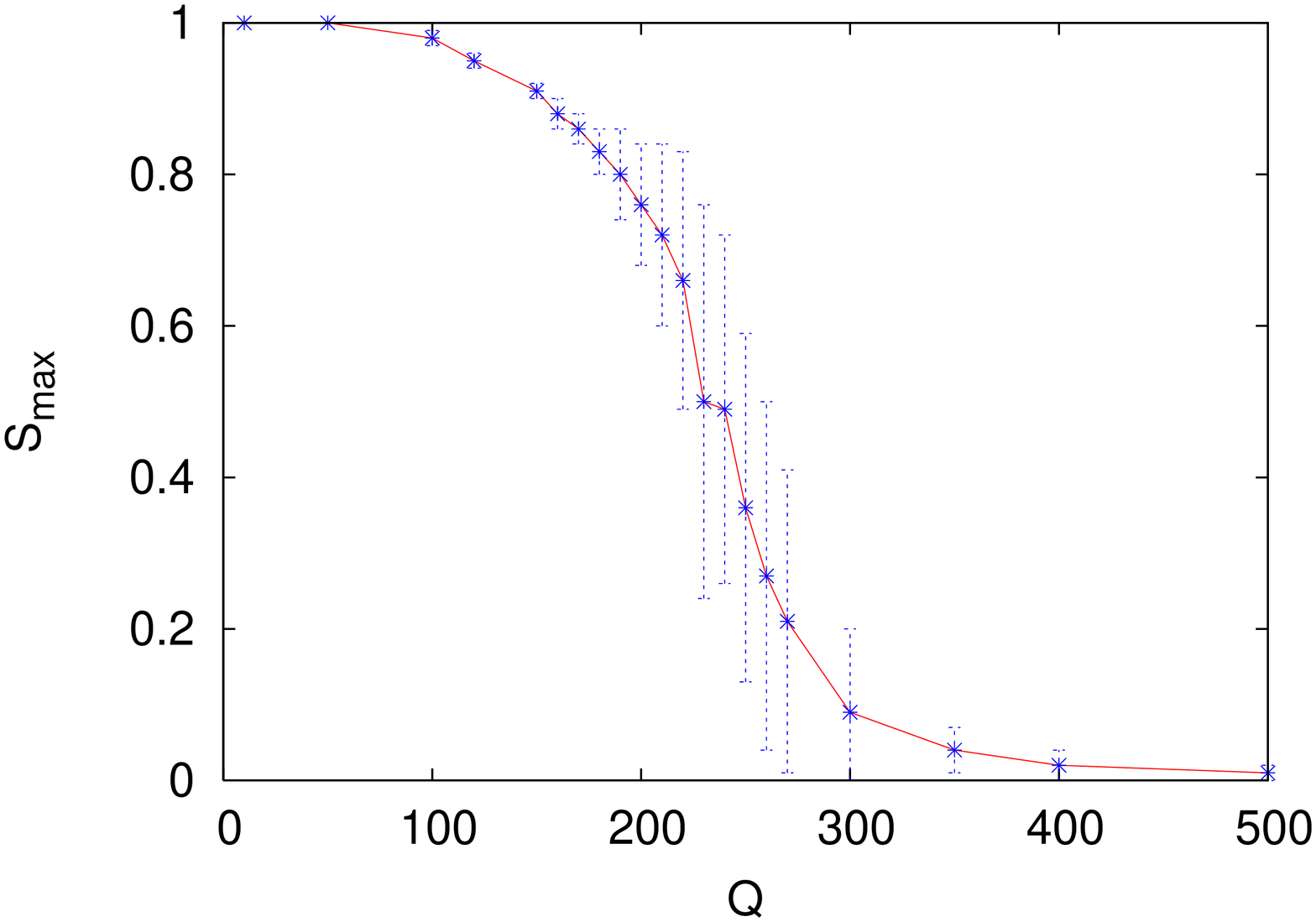,width=2.1in,angle=-90,clip=1}
\epsfig{file=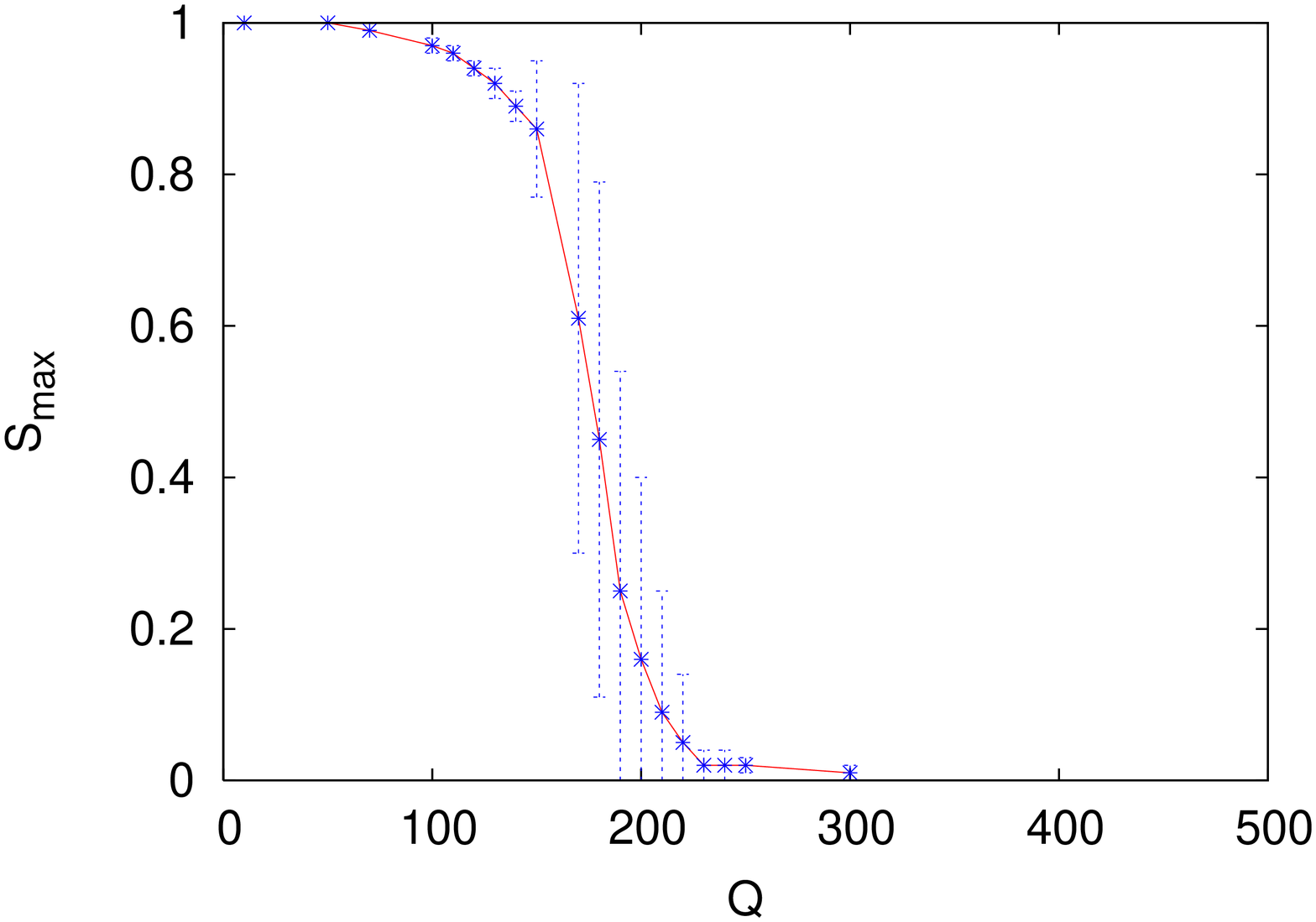,width=2.1in,angle=-90,clip=1}
\caption{(Color online) Relative size of the largest cultural
  component $S_{max}$ of the final absorbing state as a function of
  the number $Q$ of cultural traits per feature. The diagrams are
  shown for SF (top) and ER (bottom) networks for comparison. In both
  cases we use $N=10^3$, $\langle k\rangle=6$ and $F=10$. The
  statistic is performed over $100$ realizations for each value of
  $Q$.}
\label{fig:1}
\end{figure}

Iterating the above discrete-time dynamics for a number of time steps
the system reaches a state in which all the links are blocked. In
other words, the dynamics ends up in an {\em absorbing state} in which
each pair of neighboring agents have cultures that are either
identical or completely different. In general, in these absorbing
states the network is fragmented into cultural clusters of identical
individuals separated by borders composed of links having
$S_{ij}=0$. The important situation in which all the nodes of the
network are culturally identical is thus a subset of the total number,
$q^{F}$, of absorbing states. In order to characterize the final state
of the system, it is useful to measure the disorder of the
corresponding absorbing state. A useful order parameter
\cite{Castellano00} is the relative size of the largest cultural
cluster, $S_{max}$. In one extreme case of global cultural consensus,
this quantity takes the value $S_{max}=1$. In the other extreme
situation in which all individuals are totally different from their
neighbors, we have $S_{max}=1/N$.

\begin{figure}[!t]
\epsfig{file=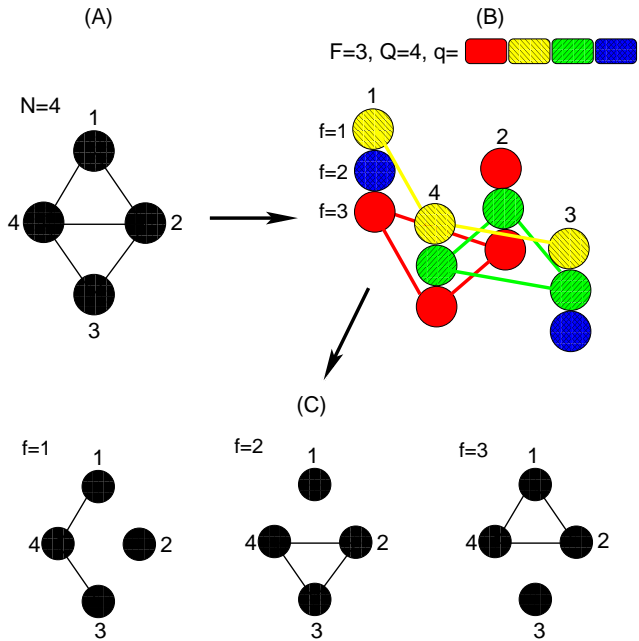,width=3.3in,angle=0,clip=1}
\caption{(Color online) Scheme of the feature level analysis. Starting
  from a network of $N=4$ nodes {\bf (A)} and an Axelrod dynamics with
  $F=3$ and $Q=4$ we stop the dynamics at some time $t$ in which the
  cultural configuration is described in {\bf (B)}. For each of the
  $F=3$ cultural features we link those nodes that share the same
  trait. Finally, we obtain $F=3$ subnetworks {\bf (C)} showing the
  consensus patterns for each cultural feature.}
\label{fig:2}
\end{figure}

According to previous studies on square lattices, when $F>2$, a
non-equilibrium first-order phase transition from order to disorder is
observed as a function of the number of traits $Q$ (the control
parameter of the phase transition) \cite{Castellano00}.  In
particular, the system always ends up in an ordered absorbing state
($S_{max}\sim 1$) when $Q<Q_c$ whereas for $Q>Q_c$ fragmented or
disordered absorbing states ($S_{max}\ll 1$) trap the dynamics of the
system. In finite complex networks, the picture is similar to the case
of square lattices, although the observed value of $Q_c$ is larger
\cite{Klemm3}.  This finding points out that network shortcuts favor
ordered states.  To show the importance of the degree of heterogeneity
of the substrate graph, we have used the model introduced in
\cite{GomezMoreno} to construct heterogeneous SF networks with
$\gamma=3$, and homogeneous Erd\"os-R\'enyi (ER) graphs ({\em i.e}
networks with $P(k)$ following a Poisson distribution \cite{ER}) with
the same average connectivity $\langle k\rangle$. In Figure
\ref{fig:1} we plot the curve $S_{max}(Q)$ for both ER and SF networks
of $N=10^3$ nodes and $\langle k\rangle=6$. In both cases we have
fixed the number of cultural features to $F=10$ and the results are
averaged over $100$ different realizations of initial conditions and
networks realizations. From the plots it is clear that in both cases
the cultural consensus decreases as the number of possible traits
increases. However, the order-disorder transition is seen to occur
more sharply in the case of ER networks where the fluctuations around
the transition region are remarkably higher than in the SF case as
shown by the error bars. Besides, it is worth remarking that it has
been found that in SF networks $Q_c(N)\sim N^{0.39}$, and thus in the
thermodynamic limit the order-disorder transition disappears and only
cultural consensus can be reached \cite{Klemm3}. Surprisingly, for
what concerns the dynamical organization towards cultural consensus, we
have not observed any qualitative difference between homogeneous and
heterogeneous architectures. For this reason, in what follows, we will
focus on the study of SF networks.

\section{Dynamical analysis at the feature level}

Besides its critical behavior, the Axelrod model in regular lattices
shows another remarkable feature: a non monotonic dynamics towards any
of its frozen equilibria \cite{Castellano00,Redner}. This non
monotonicity refers to the time evolution of the number of active
links, and is related to the fact that one interaction leading to
consensus between two agents may simultaneously destroy a higher local
consensus among one of these agents and the rest of its neighborhood
\cite{sociophysics}. This competition, due to the one-to-one nature of
the interaction, leads to extremely large time scales to reach the
final equilibrium in which all the links are blocked. Therefore, it is
interesting to study in detail how this dynamical organization takes
place in the case the model is implemented on a complex network $G$.
To this end, we analyze the formation of cultural clusters, by looking
at this process as the growth, link by link, of a network of ``{\em
  dynamical links}''. We name $G(t)$ such a dynamical network at time
$t$.  We assume that a dynamical link $i-j$ between two connected (in
$G$) individuals $i$ and $j$ is present in $G(t)$ if the two connected
individuals are culturally identical at time $t$. The dynamical
network $G(t)$ allows to characterize the transient dynamics of the
Axelrod model towards the absorbing final state by means of the time
evolution of network measures such as the number of cultural clusters
in $G(t)$, the size of the largest cultural component, or the degree
distributions of $G(t)$.  In addition to this, we will also analyze
the dynamical graph $G(t)$ as composed of $F$ different dynamical
subnetworks, $G^f(t)$, $f=1,\ldots,F$, one dynamical subnetwork for
each cultural feature.  For a given feature $f$, the graph $G^f(t)$ is
defined by considering two neighboring nodes $i$ and $j$ linked in
graph $G^f(t)$ if they share at time $t$ the same cultural trait:
$v_{i}^{f}=v_{j}^{f}$. Obviously, two neighbors can share a maximum of
$F$ links, one for each feature, meaning that the two nodes have
become culturally identical, $S_{ij}=1$. 

In Fig. \ref{fig:2} we show how the graphs $G(t)$ and $G^f(t)$ are
constructed in the case of a small graph $G$. We consider the network
$G$ with $N=4$ nodes shown in Fig. \ref{fig:2}.a), and the Axelrod
model with $F=3$ cultural features and with $Q=4$ possible cultural
traits. Suppose, at a given time $t$, the nodes of the network take
the cultural configuration shown in Fig. \ref{fig:2}.b). Then, the
dynamical graph $G(t)$ has no links, while the three graphs $G^f(t)$,
with $f=1,2,3$, are shown in (Fig. \ref{fig:2}.c). Note that no couple
of nodes in the system is culturally identical from the point of view
of the order parameter $S_{max}$. Conversely, Fig. \ref{fig:2}.c show
that a certain internal consensus is observed at the level of each of
the three features. In fact, connected components of size $3$ appear
in each of the $F=3$ subnetworks. Summing up, the analysis we propose
allows to observe the internal dynamical organization of the system by
separating the processes that occur at each single cultural feature
level, instead of integrating all the information into one single
observable.

\section{Results}

The dynamics of the Axelrod model always ends up in a frozen state, a
state in which no more changes are observed. As shown in
Fig. \ref{fig:1}, it depends on the value of $Q$ whether or not such a
frozen state corresponds to a monocultural regime. In this Section we
focus our attention on the dynamical evolution towards the frozen
state. In order to properly analyze the time evolution of a generic
quantity $X$ (such as the number of clusters, the average size of the
largest component, etc), we need to average the results over a large
number $R$ of trajectories. In particular, we have used at least
$R=50$ different realizations for each value of $Q$.  Since each
initial condition can take a different time $T_r$ ($r=1$, \ldots ,$R$)
to converge to the frozen state, the averaged evolution is obtained by
mapping the time $t$ for each realization to a common normalized time
$\tau=t/T_r$. Therefore, the corresponding averaged time evolution for
the quantity $X$ is obtained as: 
\begin{equation}
 X(\tau)=\frac{1}{R}\left[\sum_{r=1;\; t/T_{r}=\tau}^{r=R} X_{r}(t)\right]\;.
\label{eq:timenorm}
\end{equation}
Obviously, the averaged time evolution starts at $\tau=0$ and ends at
$\tau=1$.  For each initial realization we have stored the instant
configurations of the network states so that we can easily measure the
evolution of different topological quantities both at the global and
at the feature level. For each realization $r$, the instant value of
the quantity $X$ at the feature level, is denoted as $X^f_r(t)$ 
and corresponds to averaging over the values of $X$ for each of 
the $F$ features: 
\begin{equation}
 X^f_r(t)=\frac{1}{F}\sum_{j=1}^{F}X^f_{r,j}(t)\;,
\end{equation}
where $X^f_{r,j}(t)$ is the value of $X$ at time $t$, for the
realization $r$, and at the level corresponding to feature $j$
. Finally, in order to obtain the averaged time evolution $X^f(\tau)$,
we average over the $R$ different realizations by using the normalized
time $\tau$, as shown above in eq. \ref{eq:timenorm}. 

\begin{figure}[!t]
\epsfig{file=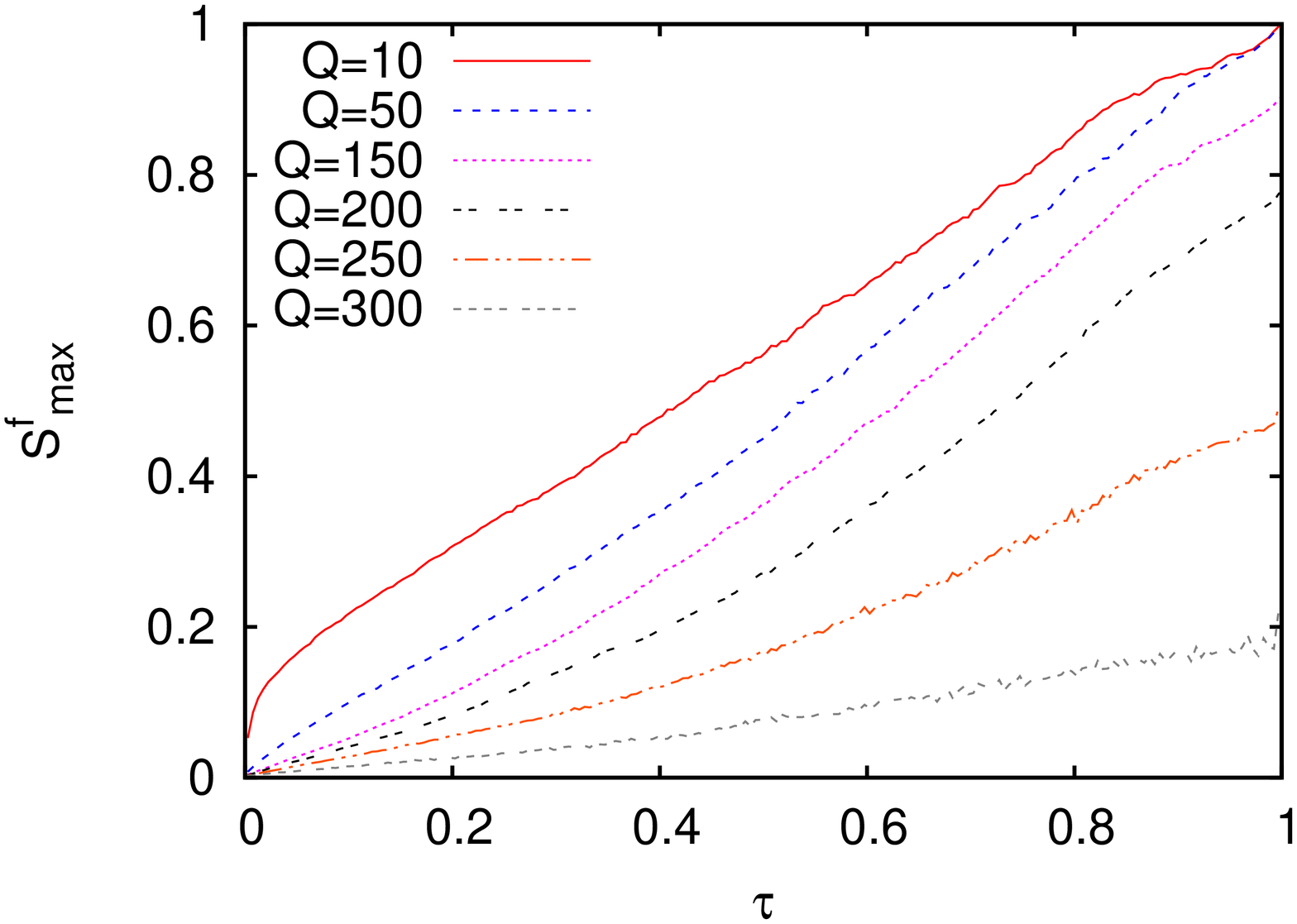,width=2.0in,angle=-90,clip=1}
\epsfig{file=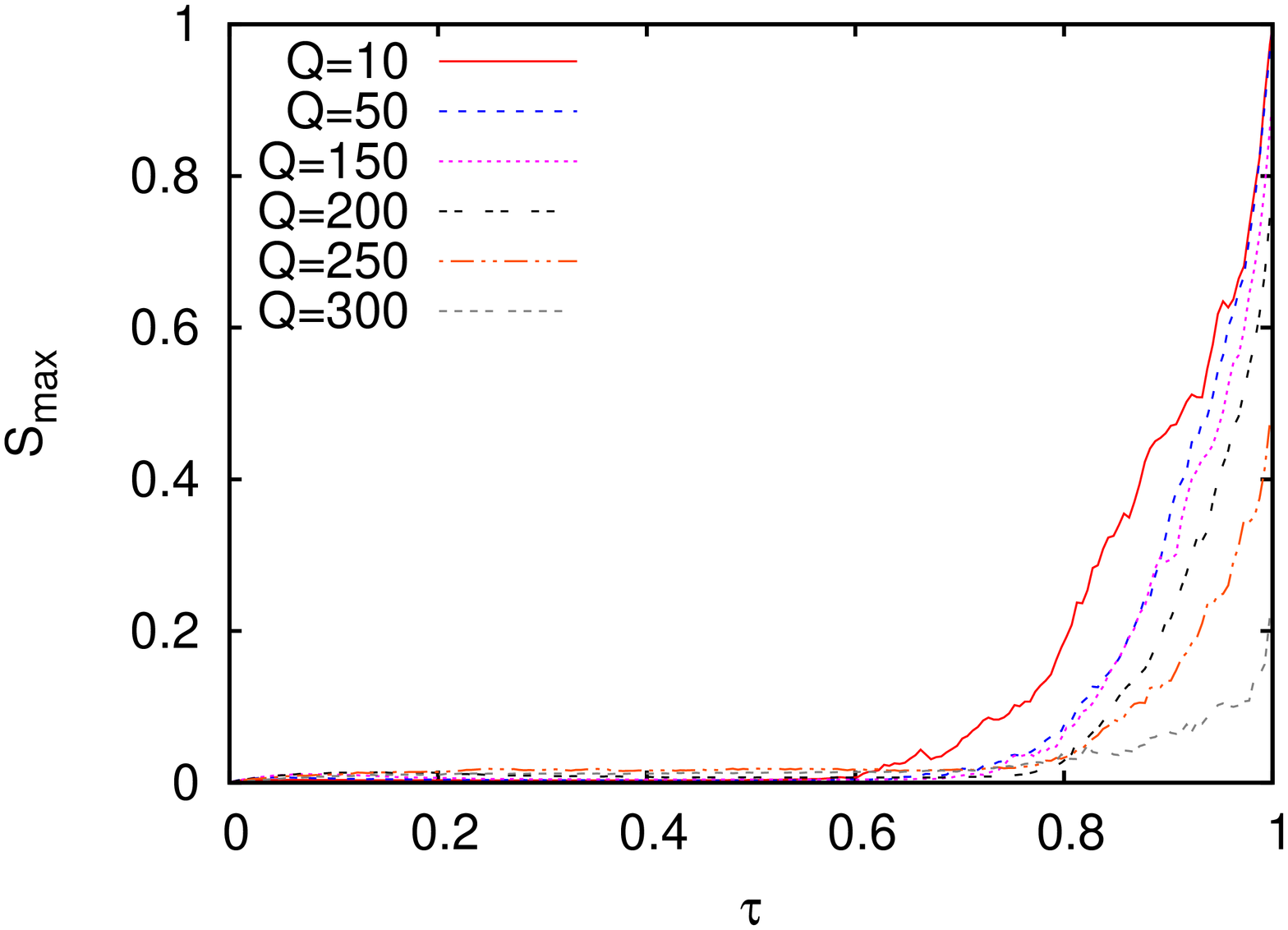,width=2.0in,angle=-90,clip=1}
\caption{(Color online) Time evolution of the size of the largest
  cultural component, $S_{max}$, at the feature (top) and at the
  global (bottom) levels.  We have considered SF networks with
  $N=10^3$ and $\langle k\rangle=6$ and $F=10$. The statistics is
  performed over $50$ different realizations.}
\label{fig:3}
\end{figure}

The first measure we consider is the size of the largest component
linking culturally identical individuals, both at the global,
$S_{max}(\tau)$, and at the feature level,
$S^{f}_{max}(\tau)$. Obviously, to compute $S^f_{max}$ we take, for
each feature, the largest component connecting individuals sharing the
same cultural trait of the corresponding cultural feature. In
Fig. \ref{fig:3} we show the time evolution of $S_{max}$ and
$S^{f}_{max}$ for several values of $Q$. The results point out that
the largest component measured at the feature level grows very fast
and that it reaches high values even when $S_{max}$ is still close to
zero. In other words, when the largest component measured at the
global scale starts growing, there are already a number of features in
which consensus has been reached, and therefore a large number of
nodes (if not all) are connected across these features. This behavior
indicates that, despite the fact that there is no signal of global
consensus, a macroscopic consensus already exist at the level of
single features.

\begin{figure}[!t]
\epsfig{file=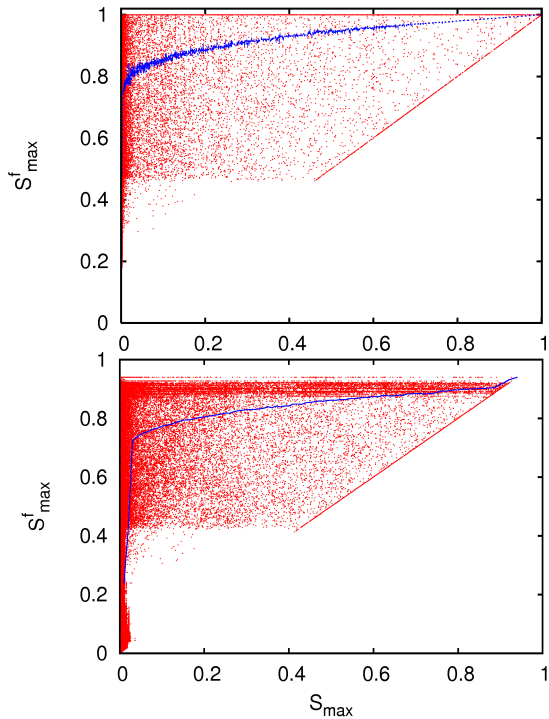,width=3.0in,angle=0,clip=1}
\caption{(Color online) The panels show the values (dots) of the
  largest cultural component of a single feature as a function of the
  size of the largest global consensus at the same time
  $(S_{max}^f)_{r,j}(t)[(S_{max})_{r}(t)]$ found during the dynamical
  evolution of the Axelrod model in SF networks with $F=10$. We
  represent the evolution of the different features ($j=1,...,F$) for
  different realizations ($r=1,...,100$).  The top and bottom panels
  correspond to $Q=20$ and $Q=150$, respectively. Both plots show that
  although most of the features have reached their own cultural
  consensus in a fast way there are few features (typically one for
  each realization) that considerably delay the global consensus of
  the system.  Finally the solid (blue) line represents the average
  size of the largest cultural component at the feature level
  $S_{max}^{f}$ as a function of the size of the largest global
  consensus component $S_{max}$.}
\label{fig:map}
\end{figure}

Figure \ref{fig:3} shows the average size of the largest cultural
clusters at the feature level by averaging across the different $F$
features and realizations. From this figure it becomes evident that
most of the features reach intra-cultural consensus in a time-scale
much faster than that observed for the global consensus. On the other
hand, the long transient dynamics towards global consensus of each
realization of the Axelrod dynamics has its roots in the
reorganization processes that occur inside each of the feature
levels. Then, how does the fast organization of the averaged feature
dynamics fit with the long transient observed at the global level? To
answer this question it is convenient to look at the evolution of each
cultural feature in a single realization of the Axelrod dynamics. In
Figure \ref{fig:map} we show a scatter plot made up with the
simultaneous values of the size of the global consensus cluster and
that for the consensus cluster inside single features at different
stages of the transient dynamics. In particular, we have plotted those
couples of values $[(S_{max})_{r},(S_{max}^f)_{r,j}]$ (where
$r=1,...,R$ and $j=1,...,F$) observed at different time steps of the
dynamics on SF networks for two different $Q$ values ($20$ and
$150$). From the plot, it is clear that at the feature level there is
a separation into fast and slow cultural features. The former features
are identified, on the one hand, by the accumulation of dots close to
$(S_{max}^f)_{r,j}=1$ ($Q=20$) and $(S_{max}^f)_{r,j}=0.9$ ($Q=150$)
and, on the other hand, by the large values of the solid (blue) curve,
$S_{max}^f(S_{max})$, constructed by averaging over the values
$(S_{max}^f)_{r,j}$ corresponding to the same value of $S_{max}$. In
addition, the slow features have their fingerprint in the accumulation
of dots along the line $S^{f}_{max}=S_{max}$ for high values of
$S_{max}$. The separation into fast and slow features is not symmetric
since, for each realization, most of the features belong to the fast
group (as pointed out by the curve $S^{f}_{max}(S_{max})$) whereas few
of them (typically $1$ as observed from the simulations) correspond to
the slow group.

\begin{figure}[!t]
\epsfig{file=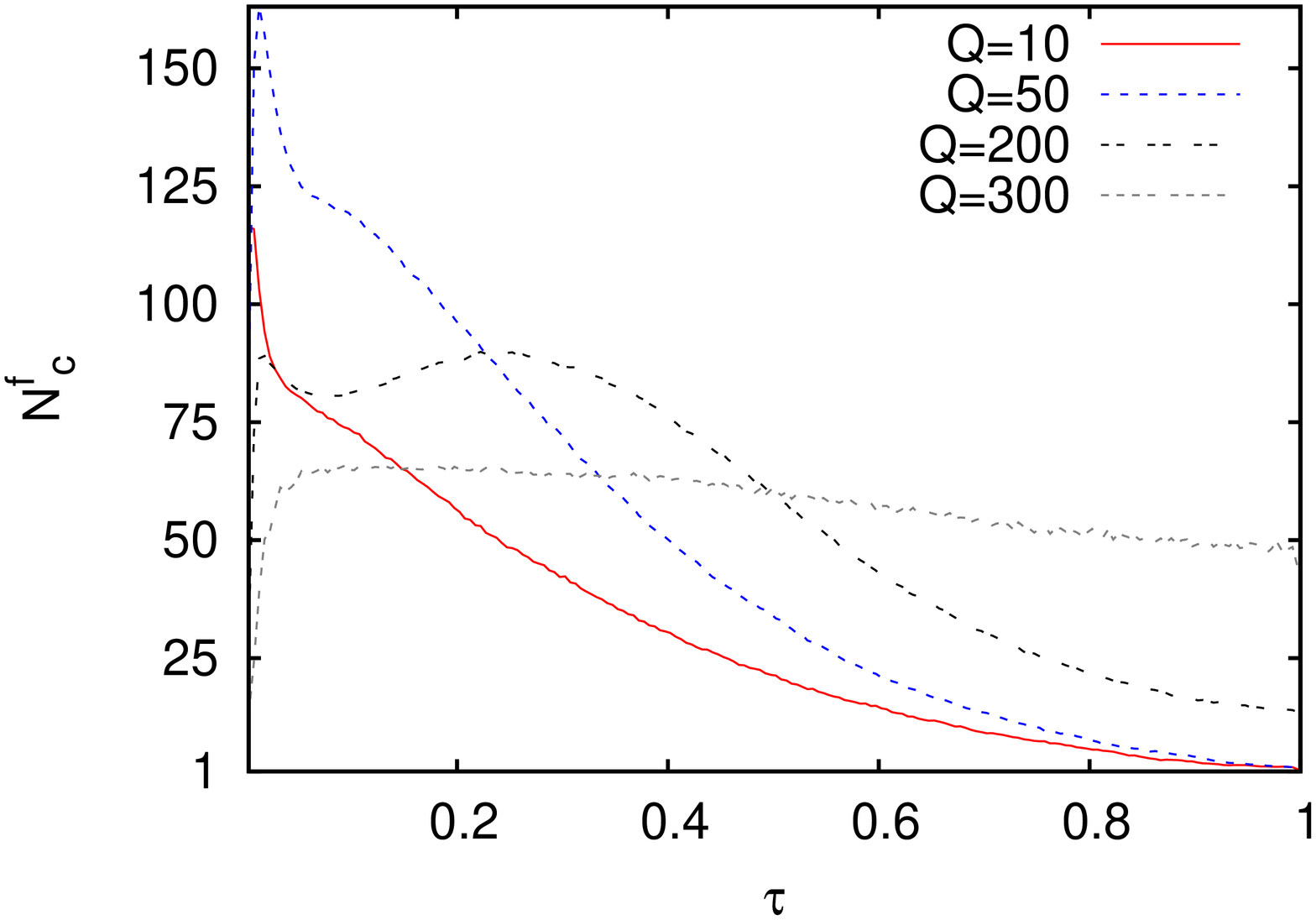,width=2.0in,angle=-90,clip=1}
\epsfig{file=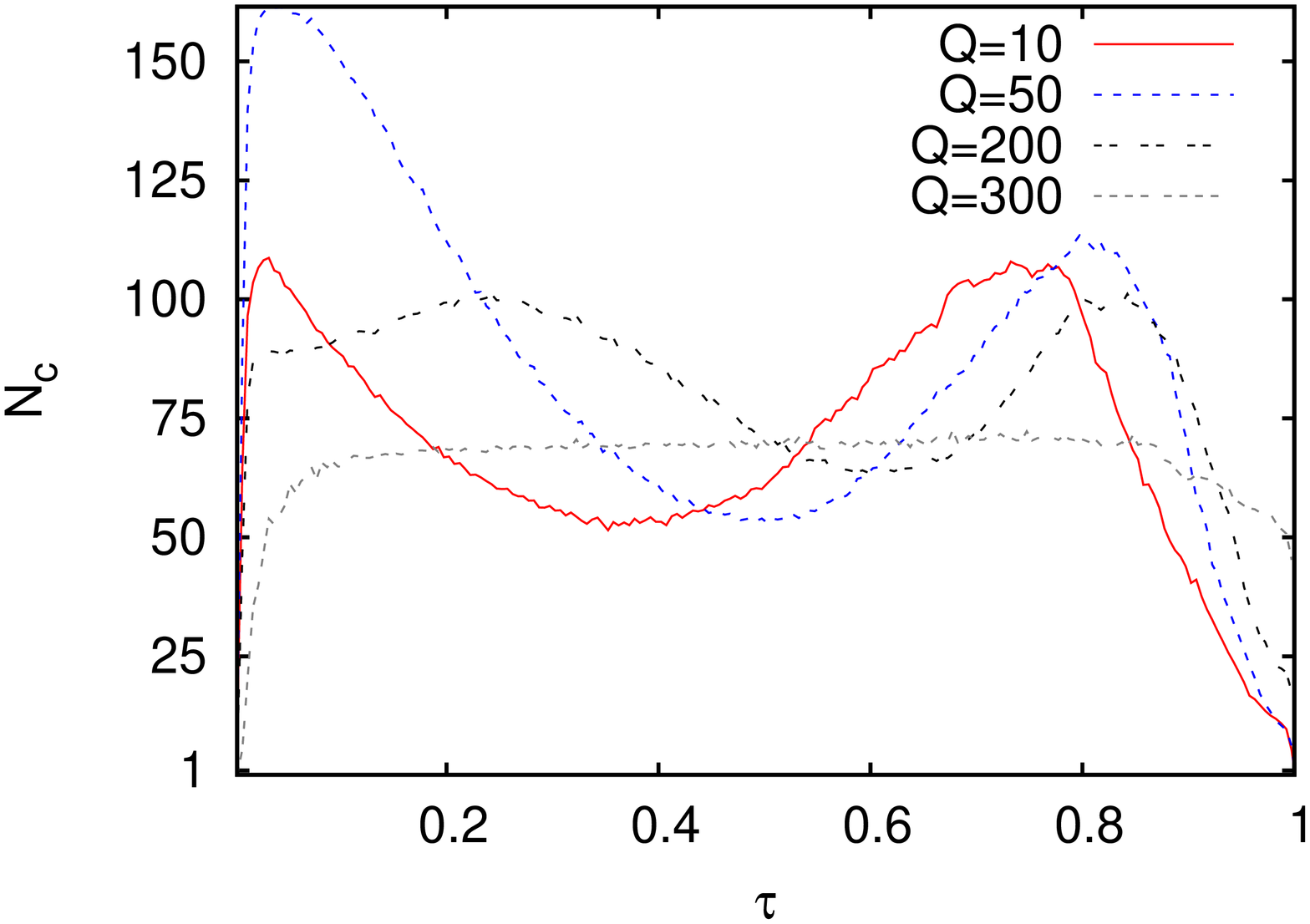,width=2.0in,angle=-90,clip=1}
\caption{(Color online) Number of monocultural clusters in SF networks
  as a function of time at the feature, $N^f_{c}$, (top) and global,
  $N_{c}$, (bottom) levels. The time has been normalized as in
  Fig. \ref{fig:3}. At the macroscopic level, the evolution of $N_{c}$
  is not monotonous while at the feature level, and for those values
  of $Q$ for which final consensus is possible, the evolution of
  $N_{c}^f$ is dominated by a decreasing pattern. The networks have
  $N=10^3$ nodes, with $\langle k\rangle=6$ and $F=10$. For the
  $N_{c}$ figure, every point shown is the average over $50$ different
  realizations, while for the $N_{c}^{f}$ one, yet an additional
  average over the $F$ features has been made.}
\label{fig:4}
\end{figure}

As for the number of monocultural clusters, we have observed that the
behavior at the feature and global levels are also distinct. Figure
\ref{fig:4} shows the results obtained for the number of clusters in
the SF network as a function of time for different values of the
parameter $Q$. At the feature level, except for short times, the
dynamics of the model leads to a monotonously decreasing number of
clusters for those values of $Q$ leading to a final equilibrium where
macroscopic consensus is reached.
Conversely, no monotonous behavior for the number of clusters is
observed at the macroscopic level. As a matter of fact, the initial
evolution of $N_c$ is close to that observed for $N_c^f$: it starts
growing for low values of $\tau$, reaches a maximum and starts to
decrease. However, in the case of $N_c$, around the transition time
from the non-consensus regime to a globally visible consensus (see
Fig. \ref{fig:3}), the number of clusters starts to grow again,
yielding a new maximum, to finally fall down very fast.
This trend of the global dynamics points out again the reorganization
taking place before reaching the final equilibrium. Once again, this
reorganization seems not to affect the equilibria at the feature
level. However, these differences between feature and global levels
are again understood when following the growth of the largest
components for each feature in single realizations of the Axelrod
dynamics. The typical evolution shows that those fast features reach
consensus in short times, $\tau\ll 1$, while those slow features
remain fragmented for a number of time steps before reaching consensus
together with the global system.
Therefore, the clear separation of time-scales at the feature level
implies that, although most of the agents reach consensus for most of
their features in a fast way, the existence of few bottleneck features
is responsible for the long transient of the global dynamics. It is
inside these slow features where the processes of reorganization occur
while the remaining (large) fraction of fast features remain unaltered
in the state of internal cultural consensus.

\begin{figure}[!t]
\epsfig{file=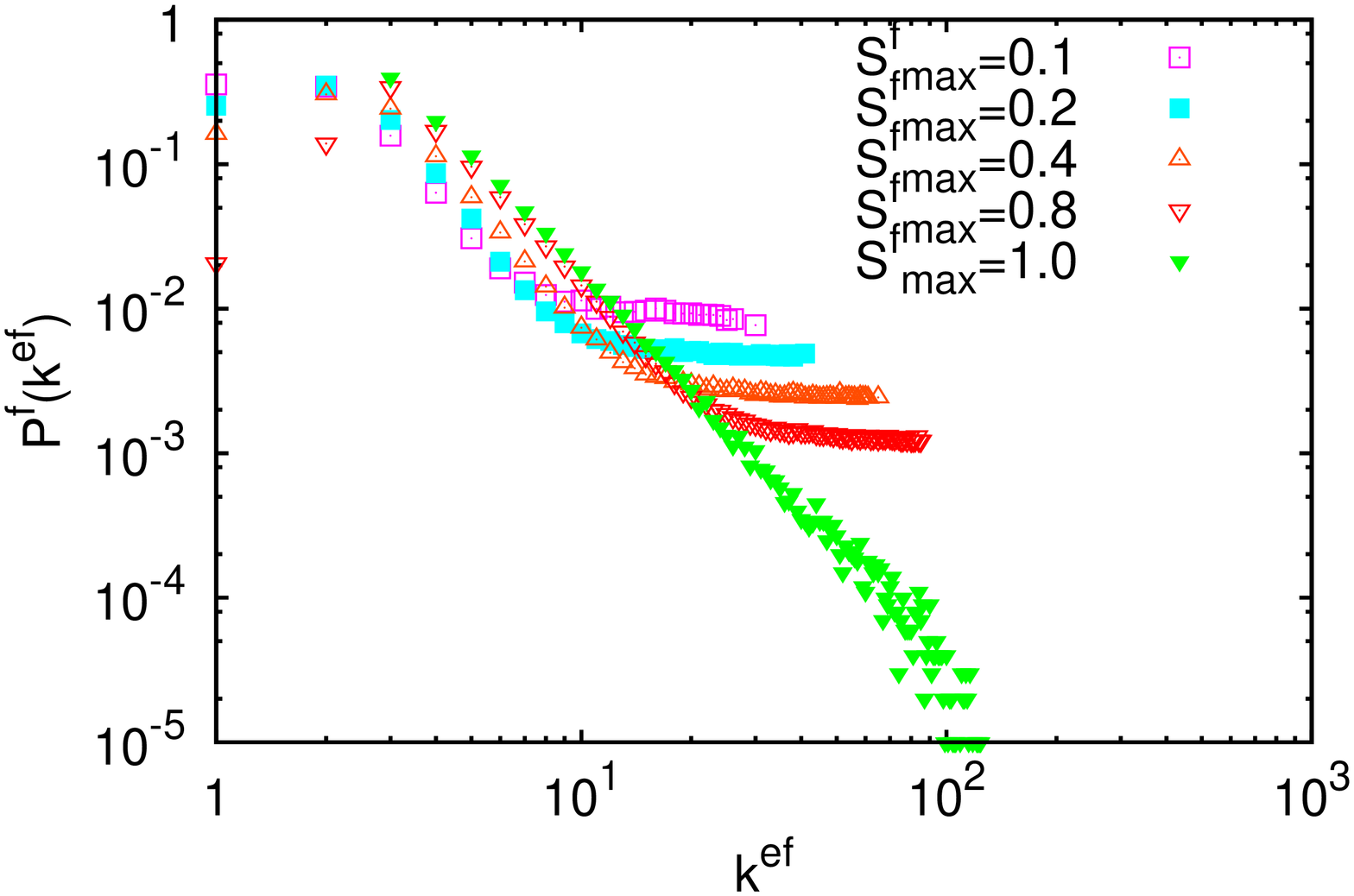,width=2.0in,angle=-90,clip=1}
\epsfig{file=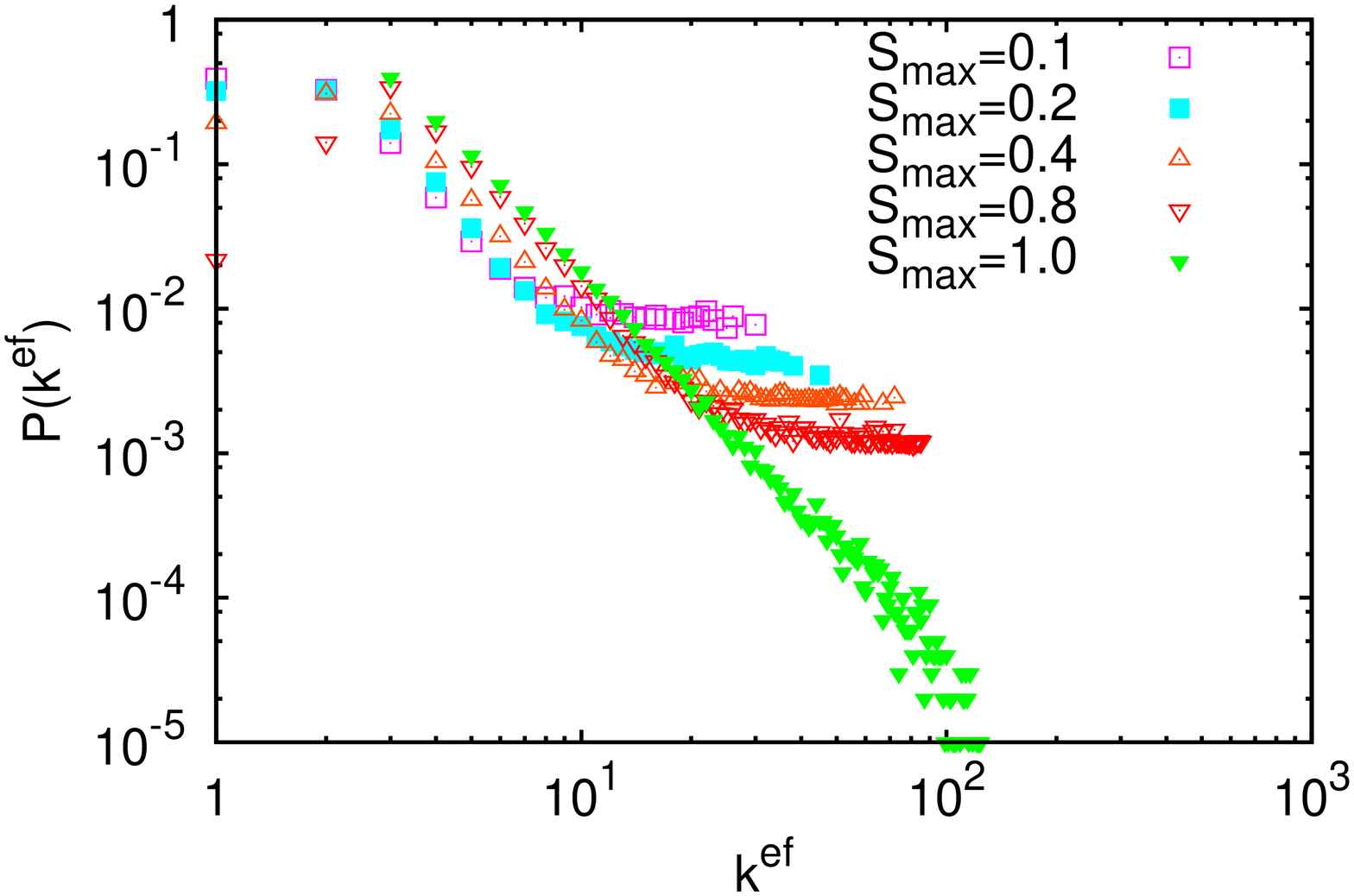,width=2.0in,angle=-90,clip=1}
\caption{(Color online) Evolution with respect to the size of the
  largest component of the effective degree distributions as measured
  in the largest monocultural clusters at both feature (top) and
  global (bottom) levels of description in SF networks. The results
  correspond to $Q=10$ and indicate that hubs are always the latest to
  reach consensus. The networks we used have $N=10^3$ nodes, with
  $\langle k\rangle=6$ and $F=10$. For the $P(k)$ figure, every point
  shown is the average over $100$ different realizations, while for
  the $P^{f}(k)$ one, yet an additional average over the $F$ features
  has been made. See the main text for further details.}
\label{fig:5}
\end{figure}

We can extract additional topological information using the approach
adopted here. Usually, consensus models define a cultural pattern when
the individuals that make up such pattern share all cultural
traits. However, it is also possible that individuals relate to each
other not because they share all the cultural features, but because
they have one or several of these traits in common. Obviously, if we
``look'' at the system at a global level, we would be unable to detect
cultural clusters unless the overlap between different individuals is
one. As seen before, however, structures also emerge at the feature
level. These ``hidden'' patterns can also be characterized
topologically.  

First we study the effective degree distribution of the giant
consensus clusters, $P(k^{ef})$. The effective degree, $k_{i}^{ef}$,
of a node $i$ belonging to a consensus cluster at the feature level is
defined as the number of its physical neighbors that share the same
cultural trait for the corresponding feature. For the global cluster
we consider only those links that join culturally identical nodes.
The effective degree distributions are normalized to the number of
nodes in the corresponding consensus components. In Fig. \ref{fig:5}
it is shown the effective degree distributions for different values of
$S_{max}$ and $S^{f}_{max}$ respectively. The number of possible
traits is set to {\bf $Q=10$} so that cultural consensus is always the
final frozen state. From the evolution of $P(k^{ef})$ as $S_{max}$ and
$S_{max}^{f}$ grow, and taking as reference the cases when
$S_{max}=S^{f}_{max}=1$, one notices that highly connected nodes are
the latest to reach consensus with its neighbors. In particular, hubs
appear overpopulating the intermediate effective degree-classes for
low values of $S_{max}$ and $S_{max}^{f}$. As the consensus components
increase so do the effective degrees of hubs and finally they reach
their physical connectivity. Interestingly, there are no significant
differences in the degree distributions measured if one follows a
feature or the system as a whole. Note, however, that the values of
$S_{max}$ are realized at different times $-$ one at the beginning
(features) and the other (global) at the end of the Axelrod dynamics.

\begin{figure}[!t]
\epsfig{file=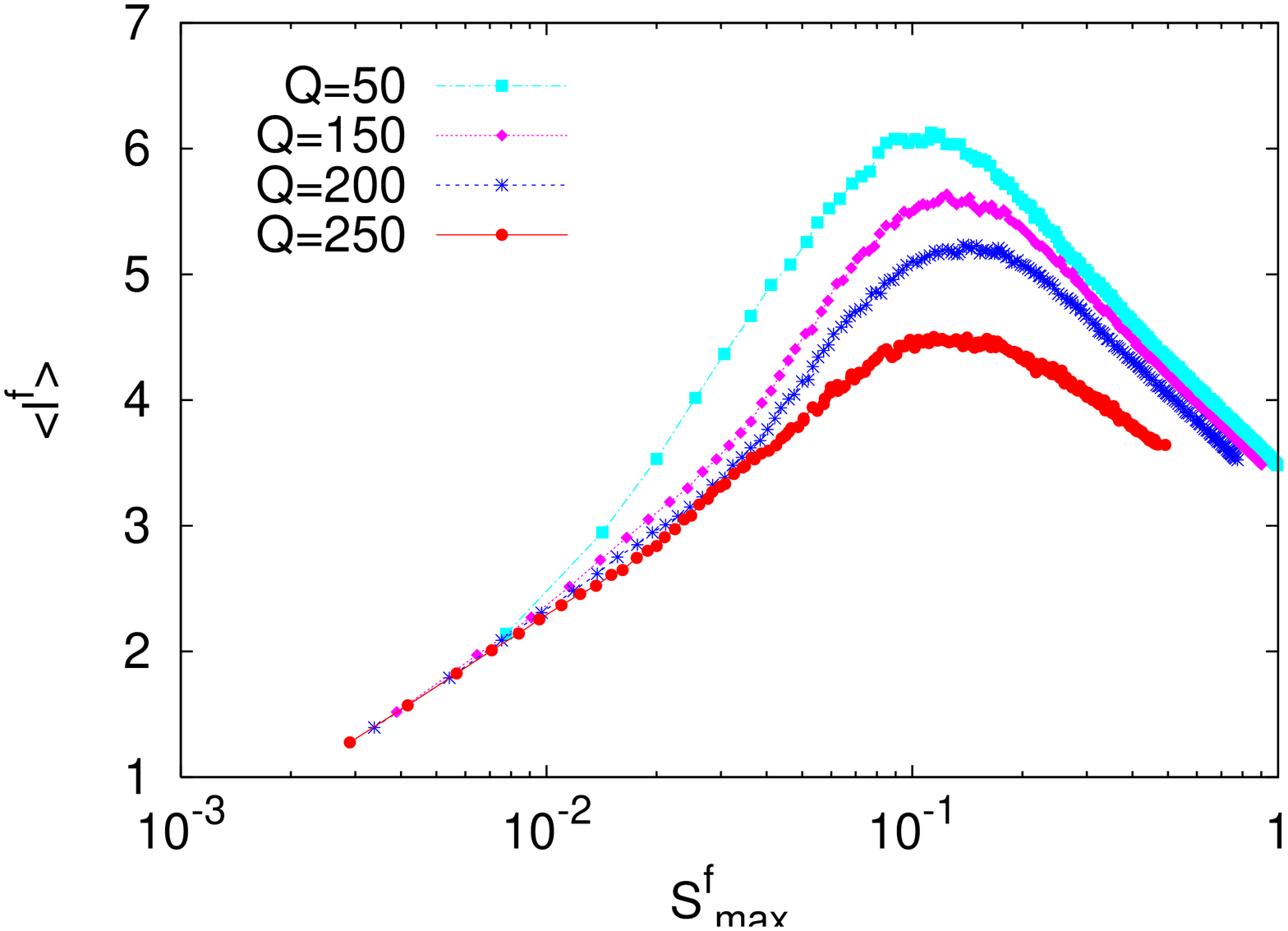,width=2.85in,angle=-0,clip=1}
\epsfig{file=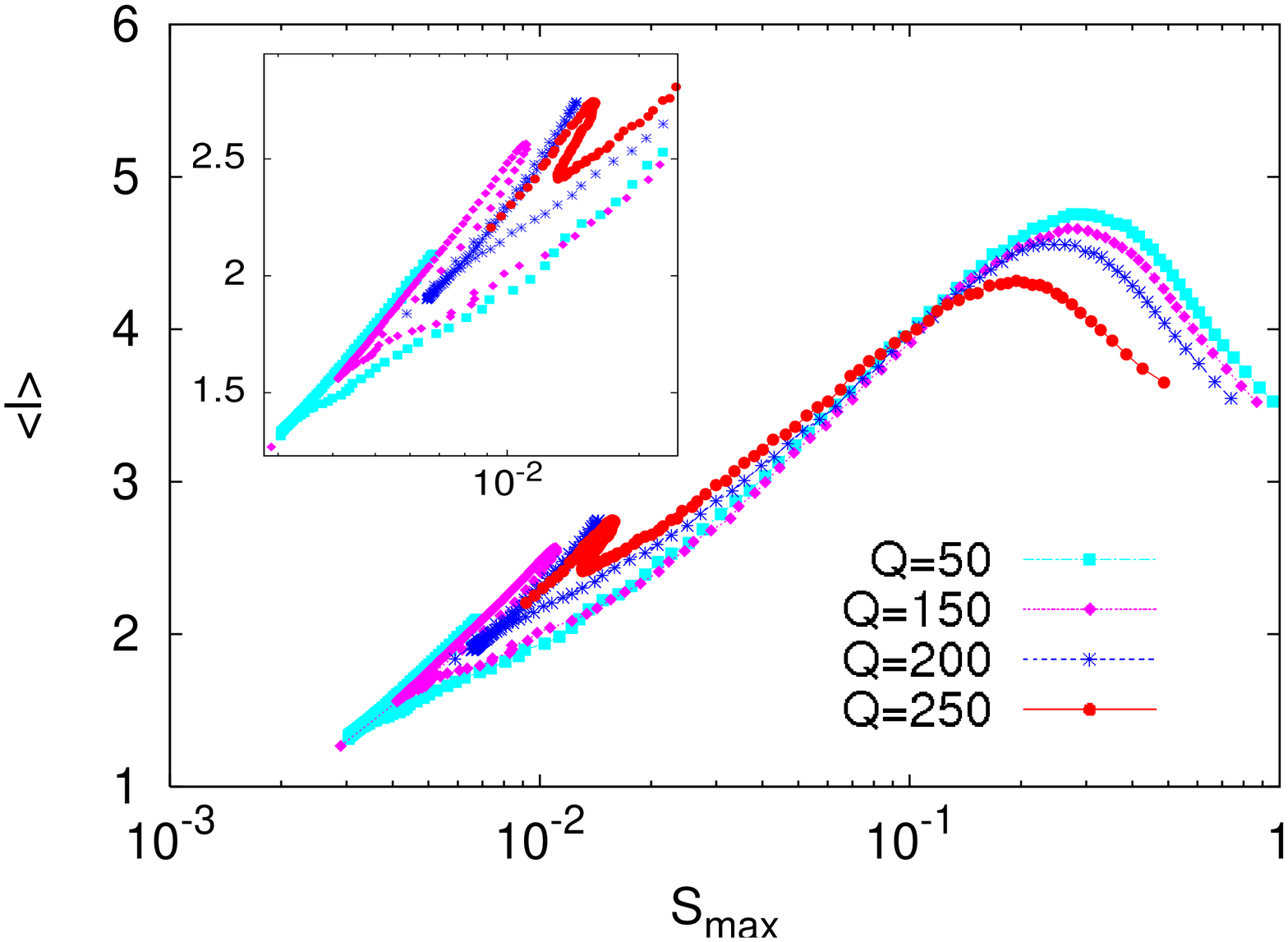,width=2.85in,angle=-0,clip=1}
\caption{(Color online) Variation of the average path length of both
  representat ions with respect to the size of the corresponding giant
  components, $S_{max}^f$ (top) and $S_{max}$ (bottom), in SF networks
  and for different values of $Q$. The networks are made up of
  $N=10^3$ nodes, with $\langle k\rangle=6$ and $F=10$. For $\langle l
  \rangle$ the statistic is performed over $50$ different
  realizations, while for the $\langle l^{f} \rangle$, yet an
  additional average over the $F$ features has been made. The inset in
  the bottom plot shows in detail the region where the curve $\langle
  l \rangle(S_{max})$ is multivalued.}
\label{fig:6}
\end{figure}

Finally, in Fig.\ref{fig:6} we show the variation of the average path
length with the size of the giant component at the feature and global
levels. The behavior of this quantity supports the phenomenological
picture previously described and allows to understand how the largest
cluster grows. For small values of $S_{max}$ and $S^{f}_{max}$, just a
few nodes distributed in many clusters have reached consensus and thus
$\langle l \rangle\approx 1$. As $S_{max}$ and $S^{f}_{max}$ grow, so
do $\langle l \rangle$ and $\langle l^{f} \rangle$, which implies that
more nodes are added to the consensus clusters, but that a significant
number of newcomers are only connected to one member of the consensus
component. In other words, when $S_{max}$ is small, the consensus
clusters (either global or at single features) grow in a tree-like
manner thus not having many loops and consequently showing large paths
connecting different nodes in these clusters. Remarkably, for small
sizes of the global consensus component, $S_{max}$, the curve $\langle
l \rangle$ presents three branches. The branch corresponding to the
largest values of $\langle l \rangle$ has its roots in the first
frustrated growth of the global consensus discussed in
Fig. \ref{fig:4}. The middle branch subsequent corresponds to the
decrease of $S_{max}$ for intermediate values of $\tau$ (pointed out
by the increase in the number of global clusters, $N_c$, in
Fig. \ref{fig:4}). Finally, the third and lowest branch of $\langle l
\rangle$ is due to the second and final growth of $S_{max}$. The large
slope of the first branch of $\langle l \rangle$ points out that the
frustrated growth of the global consensus occurs in a more tree-like
way than the final one.  Further increase of $S_{max}$ and
$S^{f}_{max}$ finally leads to an increase of the probability of
incorporating nodes sharing more than one connection with the
consensus component. Therefore, at this stage, the addition of new
nodes to the consensus components implies that a large number of links
are also incorporated into the giant consensus component and thus at
this second stage the growth is dominated by the addition of new
links.  
As for both levels of description, the difference relies on the
relative sizes of the giant consensus component at which the maxima of
$\langle l \rangle$ and $\langle l^{f} \rangle$ are attained and the
dependence with $Q$. In particular, for those values of $Q$ leading to
a final macroscopic consensus, the values of $S^{f}_{max}$
corresponding to these maxima of $\langle l^{f} \rangle$ increase with
$Q$ whereas for global consensus the trend is the opposite.
Obviously, as $Q$ grows the curves for both $\langle l \rangle$ and
$\langle l^{f} \rangle$ show the same two maxima.
This different behavior for low values of $Q$ comes again from the
existence of fast growing consensus clusters at the feature level:
While the evolution of $\langle l \rangle$ is ruled by the (few)
bottleneck features, $\langle l^f \rangle$ is mostly contributed by
those fast growing features. As a consequence, the tree-like growth
stage of the development of feature consensus clusters is replaced by
the link-dominated growth much earlier than observed at the global
level. This result points out again that the dynamical organization
depends strongly on the level of representation of the dynamics.

\section{Conclusions}

In this paper we have studied the microscopic dynamics towards
cultural consensus in the Axelrod model on SF networks. In particular,
we analysed how single traits spread across cultural
features. Comparing such microscopic dynamics with that observed at
the global level, {\em i.e.}  integrating all the features into one
cultural observable, we have shown that feature consensus is achieved
remarkably faster than global consensus. In particular, while at the
global level there are no signals of cultural consensus, most of the
cultural features have already reached a macroscopic agreement. We
have also observed important differences in the dynamic organization
towards cultural consensus. In fact, at the global level there is a
clear reorganization before cultural consensus is reached, this being
evident from the non-monotonicity in the time evolution of the number
of consensus clusters. Conversely, such reorganization processes are
localized in a few cultural features rather than taking place in all
the feature levels. Such localization points out the existence of a
fast time scale for most of the cultural features which becomes
screened when looking for the time evolution of the global consensus.

We have also analyzed the time evolution of the patterns of consensus
clusters. In clusters defined both at single feature and at global
scale, high degree, although present when macroscopic consensus is
observed, show a misrepresented effective connectivity, since not all
node neighbors are culturally identical. Additionally, the growth of a
giant consensus components has two well-differentiated stages. In the
first stage, the growth takes place in a tree-like manner, while in
the second stage the nodes attracted come along with a large amount of
links, reducing considerably the distances within these consensus
clusters. Remarkably, at the feature level the tree-like growth is
much shorter than the one dominated by the addition of links. This
points out that the development of feature consensus clusters is more
compact than its global counterpart.

Finally, it is worth stressing that we have not found any significant
difference when applying the same analysis to homogeneous ER networks
and lattice geometries, in contrast to other dynamical process where
noticeable differences in the organization of the dynamical equilibria
are observed \cite{s1,s2,Gomez-GardenesPRL,Sinatra}. This result poses
an open question on the influence of network structure on the
dynamical organization of the Axelrod model. For future work, it would
be also interesting to design strategies that favor global consensus
based on the existence of a fast time-scale for the development of
consensus at the feature level of representation.


\acknowledgements This work has been partially supported by MICINN
through Grants FIS2008-01240, FIS2009-13364-C02-01 and MTM2009-13848.


\end{document}